\documentclass[final,epsfig,amsfont]{article}
\usepackage{latexsym}
\usepackage[dvips]{graphicx}
\newcommand{\bino}[2]{\mbox{$(\stackrel{#1}{\scriptstyle #2})$}}

\begin{document}
\title{Cyclotron-Bloch dynamics of a quantum
particle in a two--dimensional lattice~II}
\author{Andrey R. Kolovsky$^{1,2}$,
Ilya Chesnokov$^{2}$, \\
$^1$Kirensky Institute of Physics, 660036 Krasnoyarsk, Russia\\
$^2$Siberian Federal University, 660041 Krasnoyarsk, Russia \\
and \\
Giorgio Mantica$^{3,4,5}$ \\
$^3$Center for Non-linear and Complex Systems,\\
Department of Science and High Technology, \\ University of Insubria, 22100
Como, Italy \\  $^4$ I.N.F.N. sezione di Milano, $^5$ CNISM unit\`a di Como.}
\date{}
\maketitle

\begin{abstract}
We study the quantum dynamics of a charged particle in a
two--dimensional lattice, subject to constant and homogeneous electric and
magnetic fields. We find that different regimes characterize these motions,
depending on a combination of conditions, corresponding to weak and strong
electric field intensities, rational or irrational directions of the electric
field with respect to the lattice, and small or large values of the
magnetic (Peierls) phase.
\end{abstract}

\section{Introduction and statement of results}
\label{sec1}

This is the second paper in a row on the quantum dynamics of a charged particle in a two-dimensional square lattice, under the influence of an in-plane electric field and a normal to the plane magnetic field, both uniform in space and constant in time. In a previous work \cite{85} we considered the case where the electric field is aligned with one of the lattice axes. We now extend this investigation to arbitrary directions of the electric field, still lying on the lattice plane. New phenomena are found in this generalization.

As discussed in \cite{85}, many physical systems, including Bose condensates in optical lattices, can be described by the Hamiltonian $H$ of a charged particle in a two-dimensional lattice oriented as the reference frame $(x,y)$,  in the tight--binding approximation:
\begin{equation}
\label{basic}
(\widehat{H} \psi)_{l,m}= -\frac{J_x}{2}\left(e^{-i2\pi\alpha m}
\psi_{l+1,m}  +  e^{i2\pi\alpha m} \psi_{l-1,m}\right)
-\frac{J_y}{2}\left(\psi_{l,m+1} + \psi_{l,m-1}\right) +ea(F_x l + F_y m) \psi_{m,l} \;,
\end{equation}
which is written here for the Landau gauge ${\bf A}=B(-y,0)$. In the above, $\psi$ is the wave-function, the integer pair $(l,m)$ labels the lattice site $(x,y)=(la,ma)$, $a$ is the lattice period, $J_x$, $J_y$ are the hopping matrix elements, $F_x$, $F_y$ are the electric field components, $e$ is the charge, and $\alpha$ is the {\em Peierls phase}, defined as the ratio between the magnetic flux through the unit cell, $Ba^2$, and the elementary flux $hc/e$.
A second ratio is crucial in this Hamiltonian, that between the Bloch frequencies $\omega_x=eaF_x/\hbar$ and $\omega_y=eaF_y/\hbar$ associated with the two components of the electric field vector. For convenience we call the ratio $\beta=\omega_x/\omega_y$ the {\em (electric) field orientation}. These two ratios deserve to be clearly marked for
further use:
\begin{equation}
\label{basic2}
\alpha:=eBa^2/hc, \;\; \beta := \omega_x/\omega_y = F_x/F_y.
\end{equation}

The content of this paper is a detailed study of the dynamics generated by the above Hamiltonian. These dynamics are far from trivial even when one of the fields is absent.  In fact, for vanishing magnetic field the wave-packet motion features Bloch oscillations, which crucially depend on the rationality of the parameter $\beta$. In the opposite case of null electric field, a commensurability condition still appears, this time in terms of the Peierls phase $\alpha$.
Clearly, the problem becomes even subtler when both fields are present. To some extent, this problem has been considered earlier in Ref. ~\cite{Naka95,Bare99,Naza01,munoz}: our goal is now to develop a systematic analysis that covers all dynamical regimes. In the paper \cite{85} we have developed this analysis in the semiclassical region $|\alpha|\ll 1$, for the particular case when the electric field is aligned with one of crystallographic axes of a square lattice: that is, we examined the case $\beta=0$. In this work we extend these studies to arbitrary directions of the electric field vector and to arbitrary magnetic field intensities. This yields a three parameters space $(F,\alpha,\beta)$ ($F$ is the amplitude of the electric field) characterized by various dynamical behaviors. We  elect not to vary a further parameter, the ratio between couplings in the two orthogonal directions of the lattice, $J_x$ and $J_y$, keeping them equal to $J$ \cite{remark1}. Also, without any loss of generality, we assume $|\alpha|\le 1/2$ and $0\le\beta \le 1$.

Our results can be summarized as follows. In sect. 3 we first focus on the case of small $\alpha$, where we use a kind of semiclassical approach in which the Peierls phase $\alpha$ plays the role of an effective Planck constant.  We derive the one-dimensional classical Hamiltonian
\begin{equation}
\label{a10}
H_{cl}=-J_y\cos P - J_x \cos Y+  {\cal F}_x P + {\cal F}_y Y \;,\quad
{\cal F}_{x,y}=\frac{ea F_{x,y}}{2 \pi \alpha}
\end{equation}
as an approximation of the Hamiltonian (\ref{basic}).  Observe that the dynamics of this system is insensitive of rationality of the parameter $\alpha$. We find that both classical and quantum dynamics strongly depends on whether the following conditions hold:
\begin{equation}
\label{basic4}
{\cal F}_x<J_x  \mbox{ and } {\cal F}_y<J_y \; .
\end{equation}
First of all, when the above are verified, a stable island gives rise to classical streaming across the lattice, in a direction orthogonal to the electric field, with velocity $v^*$ given by
\begin{equation}
\label{basic6}
  v^* = ea^2F/h \alpha .
\end{equation}
Since semiclassical theory is applicable whenever the Peierls phase $\alpha$ is much smaller than one, we therefore predict that when
\begin{equation}
\label{basic5}
eaF/2 \pi J < \alpha \ll 1
\end{equation}
the quantum motion generated by the Hamiltonian (\ref{basic}) allows for wave-packets traveling at speed
$v^*$ for any value of the direction ratio $\beta$, at least for a finite time-span that increases as $\alpha$ tends to zero. Conditions (\ref{basic5}) define the {\em small field, semiclassical regime}.

We then turn to a purely quantum analysis. In Sec.~\ref{sec3} we start by examining the case of rational orientation $\beta$ and we argue that for non-zero electric field the spectrum of (\ref{basic}) is absolutely continuous for any value of $\alpha$. We describe two techniques to compute this spectrum, that can also be implemented numerically. Using spectral analysis, when (\ref{basic5}) holds, we construct the Stark transporting states, already defined in the particular case Ref.~\cite{85}, that quantize the semiclassical transporting islands of Hamiltonian (\ref{a10}). We show that their evolution is characterized by linear motion with speed $v^*$. Also in Sec.~\ref{sec3}, we use perturbation theory to compute the leading order term, in the inverse field amplitude $1/F$, of the width of the energy bands of Hamiltonian (\ref{basic}). Letting $\beta=r/q$, with $r$,$q$ co-prime integers, we find that these widths scale as $(1/F)^{q+r-1}$. This suggests that the energy spectrum of this Hamiltonian is {\em pure point} for irrational direction ratio $\beta$.  Note that being based on quantum perturbation theory, these results are not restricted to the semiclassical region.

Next, we turn to the numerical simulation of the time evolution of an initial wave-packet, to confirm the theoretical predictions of the spectral analysis. The numerical techniques are briefly described in Sec.~\ref{sec4}. They permit to attack the most general case, that includes non-rational directions of the electric field. First we consider the case $F\gg F_{cr}\equiv 2\pi\alpha J/ea$, where we can apply the perturbation theory of Sec.~\ref{sec3}, and show that an initial wave-packet spreads in the direction orthogonal to the electric field with a dispersion $\sigma$ that increases in time as
\begin{equation}
\label{basic7}
\sigma \sim (1/F)^{(q+r-1)} t
\end{equation}
(here, as above, $\beta=r/q$). This behavior sets in after a transient time,
\begin{equation}
\label{basic8}
\tau \sim F^{(q+r-1)} ,
\end{equation}
during which the packet is localized. Thus for irrational field directions the wave packet is always localized in the lattice, which is consistent with discreteness of the spectrum for irrational $\beta$.

Next, in Sec.~\ref{secsemevo} we describe the case $F<F_{cr}$ for the semiclassical region $|\alpha| \ll 1$. We find that the motion is either a ballistic spreading (for an incoherent initial wave-packet) or directed transport (for a properly devised coherent initial wave-packet), both for rational and irrational directions $\beta$, for all times that we could reach in our numerical simulations. This seems to be inconsistent with discreteness of the spectrum for irrational $\beta$. To resolve this seemingly contradiction we turn in Sec.~\ref{secgen} to the case of arbitrary values of the Peierls phase $\alpha$. We simulate the quantum evolution for an increasing sequence of values of $\alpha$ while keeping the classical parameters ${\cal F}_x$ and ${\cal F}_y$ fixed. For irrational directions $\beta$, at fixed time, we observe a sharp suppression of ballistic spreading, occurring when $\alpha$ overcomes a certain threshold. Turning then to a time-resolved analysis, we observe saturation of the ballistic spreading, for times larger than a threshold that grows abruptly with the inverse of the electric field intensity $F$. Combined with the observation on band widths, this fact strengthen the conjecture that the spectrum of the system is pure point, for any irrational direction $\beta$; yet the localization length should depend sensitively on the magnitude of the electric field $F$, like in certain two-dimensional models of quantum rotators and of Anderson localization \cite{bellima,belli1,belli2}.

\section{Aligned electric field: review of results}
\label{sec2}

To introduce concepts and notations, we now briefly review the results of our previous work \cite{85}, where we considered the case of an electric field ${\bf F}=(0,F)$ aligned with the $y$ axis of a reference frame defined by a two-dimensional square lattice. The magnetic field is directed orthogonally to
the lattice plane, and the Peierls phase $\alpha$ is small. The system shows two qualitatively different dynamical regimes, depending on the magnitude of the electric field: a regime of directed transport for weak fields and a regime of ballistic spreading for strong fields, separated by a critical magnitude, $F_{cr}=2\pi\alpha J_x/ea$.  Although we will keep track of physical constants for completeness in what
follows, in numerical experiments we adopt adimensional units so that $e=a=\hbar = 1$ and we set $J_x=J_y=1$. Also, until Sec.~\ref{secgen} or unless otherwise noted, the Peierls phase is $\alpha = 1/10$.

\subsection{The energy spectrum}

\begin{figure}
\center
\includegraphics[width=10cm,clip]{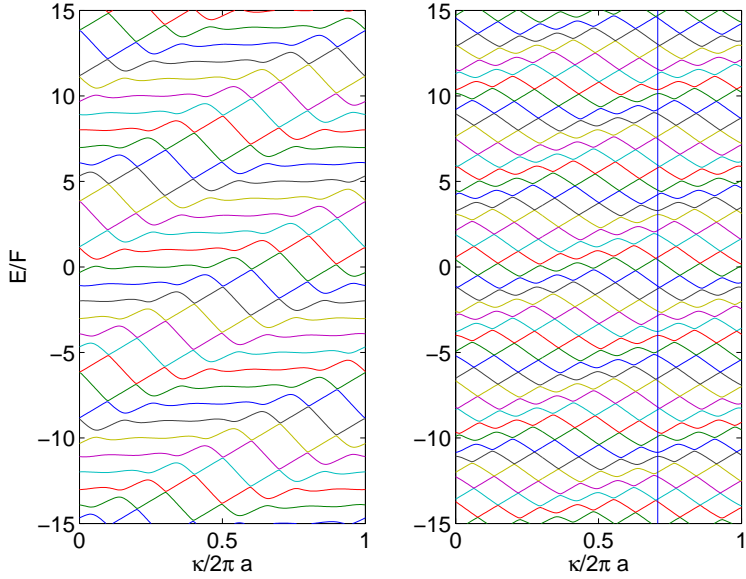}
\caption{(Color online) The energy spectrum $E=E_\nu(\kappa)$ for $F=0.3$ and $(r,q)=(0,1)$ (left) and $(r,q)=(1,1)$ (right).  The vertical line in the right panel marks the first Brillouin zone, the size of which scales as $1/\sqrt{r^2+q^2}$.}
\label{fig1}
\end{figure}

The stationary Schr\"odinger equation derived from Eq. (\ref{basic}) is
\begin{equation}
\label{1}
 -\frac{J_x}{2}\left(e^{-i2\pi\alpha m} \psi_{l+1,m}  +
e^{i2\pi\alpha m} \psi_{l-1,m}\right) -\frac{J_y}{2}\left(\psi_{l,m+1} +
\psi_{l,m-1}\right) +eaFm \psi_{l,m} = E \psi_{l,m} \;.
\label{oned1}
\end{equation}
Imposing a periodicity of period $L_x = L a$ in the $l-$direction in Eq. (\ref{1}), where $L$ eventually tends to infinity, and using the substitution
$\psi_{l,m}= \frac{e^{i\kappa al}}{\sqrt{L}} b_m(\kappa),$
where $\kappa$ is the dimensional quasimomentum, $\kappa=2\pi k/a L$,  with $k=0,1,\ldots,L-1$,  taking values in $[0,2\pi/a)$, one reduces (\ref{1}) to the following 1D equation for the coefficients $b_m(\kappa)$:
\begin{equation}
\label{3}
-\frac{J_y}{2}(b_{m+1}+b_{m-1}) - J_x\cos(2\pi\alpha m-a\kappa)b_m +
eaFmb_m =E b_m \;.
\end{equation}
The spectrum of (\ref{3}) for a fixed $\kappa$ is a modulated Wannier-Stark ladder and the energy bands appear by varying the quasimomentum. For example and for the purpose of future comparison, the left panel of Fig. \ref{fig1} shows the eigenvalues $E_\nu(\kappa)$ of this equation for $F=0.3$.  To understand the displayed spectrum the semiclassical approach proves to be very useful.

\subsection{Semiclassical theory}
\label{sec2c}

\begin{figure}
\center
\includegraphics[width=10cm,clip]{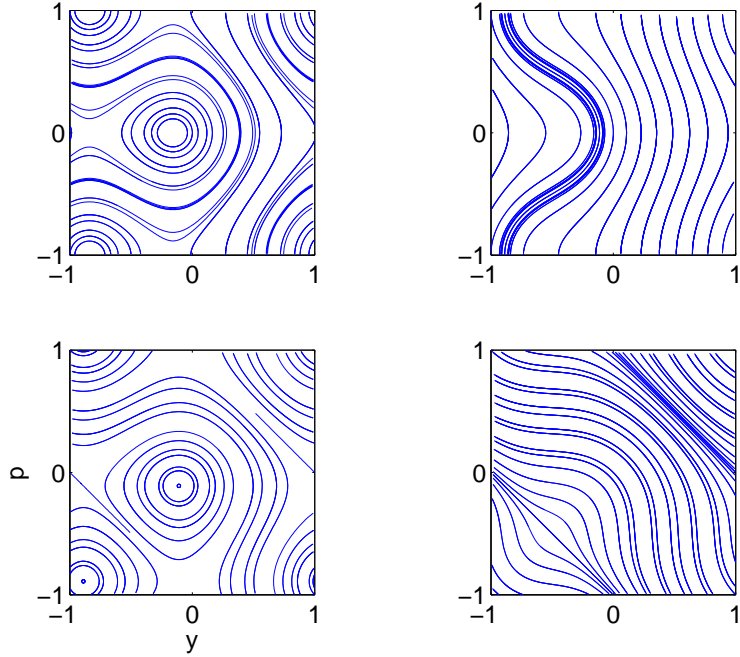}
\caption{Phase portrait of the classical system (\ref{a1}) for $(r,q)=(0,1)$
(upper row) and $(r,q)=(1,1)$ (lower row) at electric field intensities
$F=0.3$ (left column) and $F=1$ (right column). Twenty different trajectories
are shown in each case.} \label{fig4}
\end{figure}

When $|\alpha|$ is much less than one, the period of the oscillating phase in Eq. (\ref{3}) is much larger than the lattice period, so that we can approximate the discrete function $b_m$ by a continuous function ${\cal B}$ of the coordinate  $y$. Furthermore, consider the shift operator $\widehat{T}_a = \exp(a \partial_y)$, that acts as $\widehat{T}_a {\cal B}(y) = {\cal B}(y+a)$. Using this in Eq.~(\ref{3}) leads to
\begin{displaymath}
 [-J_y \cos(i a \partial_y) -J_x \cos(2 \pi \alpha  y/a - a \kappa) + eF y] {\cal B}(y) = E {\cal B}(y).
 \end{displaymath}
Next, introducing the operators $\widehat{Y}=2 \pi \alpha y / a$ and $\widehat{P}=-ia \partial_y$ in the above leads us to the effective  Hamiltonian
$\widehat{H}_{qu}= -J_y\cos \widehat{P} - J_x \cos( \widehat{Y}-a\kappa) + {\cal F} \widehat{Y}$, whose classical counterpart reads
\begin{equation}
\label{8}
H_{cl}= -J_y\cos P - J_x \cos Y + {\cal F} Y \;, \quad
{\cal F}=\frac{ea F}{2 \pi \alpha} \;.
 \label{eqoned2}
\end{equation}
Since in the quantum description the canonical variables $P,Y$ are operators obeying the commutation relation $[\hat{Y},\hat{P}]=i2\pi\alpha$, the semiclassical parameter is here the Peierls phase $\alpha$. If $|\alpha | \ll 1$, one can explain certain features of the spectrum in Fig.~1(a) by simply analyzing the phase portraits of the classical system (\ref{8}). In particular, it is easy to see that the Hamiltonian (\ref{8}) can support bounded motions if $F<F_{cr}$: these are seen as the islands around the fixed points in Fig.~\ref{fig4}, that draws the solutions of Hamilton equations for the Hamiltonian $H_{cl}$, 
Eq.~(\ref{8}).

By introducing the Bloch frequency  $\omega=eaF/\hbar$ and using the canonical substitution $P\rightarrow P-\omega t$, one obtains from (\ref{8}) the time-dependent Hamiltonian
\begin{equation}
\label{9} H_{cl}(t)= -J'_y\cos ( P - \omega t) - J'_x \cos Y  \;,\quad
J'_{x,y}=2\pi\alpha J_{x,y} \;,
\end{equation}
in which the canonical variables lie in the torus, $-\pi \le P,Y <\pi$. The classical phase space of solutions of (\ref{8}), pictured in Fig. \ref{fig4}, can also be calculated as the stroboscopic map of the time--periodic Hamiltonian (\ref{9}). Two cases for $F<F_{cr}$ and $F>F_{cr}$ are shown in Fig.~\ref{fig4}(a,b). Bounded trajectories of (\ref{8}) appear now as two nonlinear resonances (transporting islands), whose size shrinks to zero when $F=F_{cr}$. Quantizing these islands leads to the
transporting modes, defined below.

Note that in Ref.~\cite{85} a different gauge, ${\bf A}=B(0,x)$, was employed. Needless to say, the quantum spectrum of the system does not depend on the gauge adopted. In classical mechanics, the latter gauge corresponds to the Hamiltonian $\bar{H}_{cl}(t)= -J'_x\cos P  - J'_y \cos(X+\omega t)$, that can be obtained from (\ref{9}) by a canonical transformation that interchanges the
role of $P$ as momentum and $Y$ as coordinate.

\subsection{Directed transport regime}
\label{sec2b}

When $F$ is small, the energy bands $E(\kappa)$ overlap forming a rather
complicated band pattern. If $F<F_{cr}$ this pattern contains a number of
parallel straight lines, seen in Fig.~\ref{fig1}(a). In the semiclassical
approach these lines are associated with nonlinear resonances, that transport
the particle in the $x$ direction at the drift velocity (\ref{basic6}).
The explicit form of the quantum transporting states is
\begin{equation}
\label{6} \Psi_{l,m}=\int g(\kappa) e^{i\kappa al} b_m(\kappa) {\rm d}\kappa
\;,
\end{equation}
where ${\bf b}(\kappa)$ is the eigenvector of (\ref{3}) associated with a
given straight line in the energy spectrum and $g(k)$ is an arbitrary square
integrable function of the quasimomentum. Note that in ${\bf
b}(\kappa)$ we jump over avoided crossings when following a straight line in
$E(\kappa)$: this is the so-called diabatic approximation. Thus the vector
function ${\bf b}(\kappa)$ in Eq.~(\ref{6}), which implicitly assumes the
extended Brillouine zone picture, satisfies the translational relation
\begin{displaymath}
b_m(\kappa+2\pi/a)\approx b_{m+1/\alpha}(\kappa) \;,
\end{displaymath}
which is exact if $1/\alpha$ is an integer. This relation should be
compared with the relation $b_m(\kappa+2\pi/a)=b_{m}(\kappa)$ which holds when
we follow the energy bands adiabatically.
\begin{figure}
\center
\includegraphics[width=10cm,clip]{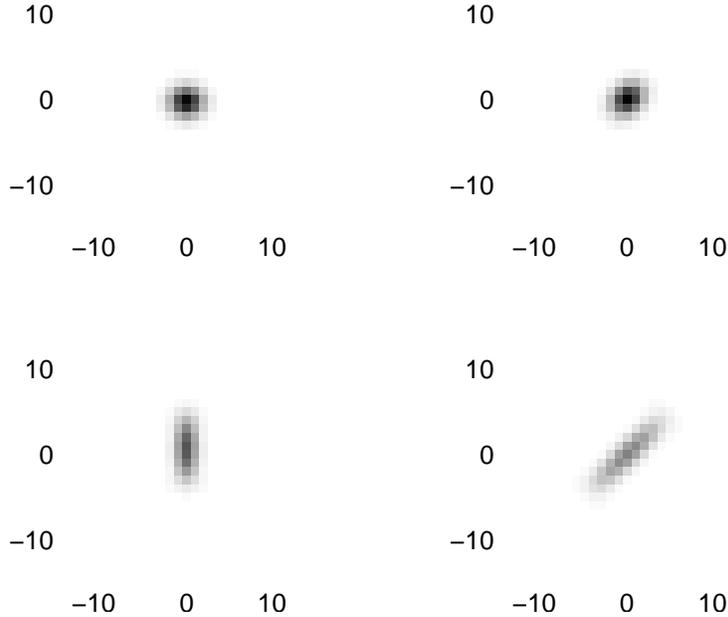}
\caption{Example of transporting states $|\Psi_{l,m}|^2$ for $(r,q)=(0,1)$
(left column) and $(r,q)=(1,1)$ (right column) depicted as grey tone images
(black maximum, linear intensity scale). The upper row is characterized by the
width parameter $C=1$, while the lower row has $C=0.2$. The other
parameters are $\alpha=0.1$ and $F=0.1$.}
\label{fig5}
\end{figure}

The left row in Fig.~\ref{fig5} shows transporting states for
$g(\kappa)\sim \exp[-C (a\kappa/2\pi)^2]$ for two different values of the
parameter $C$  \cite{remark2}. Since the slope of the straight
lines in the spectrum is equal to the drift velocity, the time evolution of
these states is a shift of the wave packet in the positive direction at
velocity $v^*$.  Thus the wave-packet width $\sigma$ is constant in time,
while the first moment grows as $M_1(t)=v^*t$. Here and below we identify the
wave-packet width with the dispersion, i.e., $\sigma=\sqrt{M_2-M_1^2}$, where
$M_{i}, i =1,2$ are the first two moments of the position operator.

It is worth stressing that the above statement refers only to initial
conditions given in Eq.~(\ref{6}).  For a generic initial wave packet one
observes an asymmetric ballistic spreading (see Fig.~10 in Ref.~\cite{85}),
where $M_1(t)\approx 0$ and $\sigma(t)\rightarrow At$. The coefficient $A$ is
here defined by the drift velocity: $A\sim v^*\sim F$ and {\em not} by the
hopping matrix element $J_x$ as in the strong field regime described
in the next subsection.

\subsection{Ballistic spreading regime}
\label{sec2a}

Consider the strong field limit, $eaF\gg J_{x,y}$. Using first order
perturbation theory the solution of (\ref{3}) reads
\begin{eqnarray}
\nonumber
E_\nu(\kappa)=eaF\nu-J_x\cos(a\kappa+2\pi\alpha\nu) \;, \\
\nonumber
b_m^{(\nu)}(\kappa)=\delta_{\nu,m} \pm (J_y/eaF)\delta_{\nu,m\pm1}
\;.
\end{eqnarray}

It follows from the above equations that the time evolution of a generic
localized wave packet is a ballistic spreading in the positive and negative
$x$ directions, so that the packet width $\sigma$ grows asymptotically as
$\sigma(t)= At$.  The numerical factor $A$ in this asymptotic expression
depends on the particular form of the initial wave packet, yet it is
superiorly bounded by $A=J_x/\sqrt{2}$. The maximal value is reached for an
incoherent wave packet, that we can mimic by assigning  random phases to the
complex amplitudes of an initial gaussian wave packet. Further down in this
paper, Fig.~\ref{fig2lin} display the $F$-dependence of the
coefficient $A$, obtained by direct numerical simulation of the system
dynamics, versus $F=F_y$. The coefficient $A$ tends to zero in the limit of
weak electric fields, while in the opposite limit of strong fields it
approaches the constant value $A=J_x/\sqrt{2}$. We will now investigate how
this situation changes when the electric field is not aligned with one of the
axes of the crystal.

\section{General direction of the electric field: Semiclassical theory}
 \label{sec3a}

The classical Hamiltonian (\ref{a10}) for generic direction of the electric field,
\begin{equation}
\label{a1}
H_{cl}=-J_y\cos P - J_x \cos Y+ {\cal F}_y Y+ {\cal F}_x P \;,\quad
{\cal F}_{x,y}=\frac{ea F_{x,y}}{2 \pi \alpha} \;,
\end{equation}
can be guessed a straightforward generalization of the classical Hamiltonian (\ref{8}).
More rigorously, the same expression can be obtained as done in Appendix, starting from the quantum equation (\ref{b6}) below. Again, it is convenient to use the canonical transformation $P\rightarrow P-\omega_y t$ and $Y\rightarrow Y+\omega_x t$, which leads to the time-dependent Hamiltonian
\begin{equation}
\label{a2}
H_{cl}(t)= -J'_y\cos(P - \omega_y t) - J'_x \cos(Y+\omega_x t)  \;,\quad
J'_{x,y}=2\pi\alpha J_{x,y} \;,
\end{equation}
Now the phase space of the system (\ref{a2}) can be reduced to the torus only when  two Bloch frequencies $\omega_x$ and $\omega_y$ are commensurate, i.e., if
$\beta:=\omega_x/\omega_y=F_x/F_y=r/q$ ($r,q$ are co-prime integers). However,
semi-classical analysis can be applied equally well in the both cases of rational and irrational $\beta$. Let us also note that  the system (\ref{a2}) has the global integral of the motion,
\begin{displaymath}
{\cal I}=H_{cl}(t)+{\cal F}_y Y +{\cal F}_x P  \;,
\end{displaymath}
that reflects the fact that the original Hamiltonian Eq.~(\ref{a10}) is one-dimensional and hence is trivially integrable, for any value of the ratio $\beta$.

The lower panels in Fig.~\ref{fig4} show the classical phase space for $\beta=1$: classical islands appear for small field magnitudes. The structure of phase space changes when the field magnitude is increased, similar to what happens in the case $\beta=0$ (higher panels). The transporting islands disappear when at least one of the two conditions, ${\cal F}_x<J_y$ and ${\cal F}_y<J_y$ is violated: this yields
the condition (\ref{basic4}). This result indicates that the original quantum system must have two qualitatively different regimes, which are separated by a critical field magnitude $F_{cr}=2\pi\alpha J/ea$. This conjecture is supported by analysis of the energy spectrum of the quantum system, that we consider now, starting from the case of rational orientation $\beta$.

\section{Energy spectrum for rational orientation of the electric field}
\label{sec3}

We now proceed to the case of arbitrary, yet rational, direction of the vector
${\bf F}=(F_x, F_y)$.  We first review two techniques to compute the energy
spectrum in the case of rational ratio $\beta = F_x/F_y$. This analysis
facilitates the understanding of the wave-packet dynamics, that will be
described in Sec.~\ref{sec4}. Letting $r$ and $q$ be relatively prime integer
numbers, we align the electric field with the $(r,q)$ direction in the plane:
\begin{equation}
\label{field} {\bf F}= \frac{F}{\sqrt{N}}\left( r,q \right) \;
\end{equation}
where $N=r^2+q^2$. We choose the gauge
\begin{equation}
\label{b2} {\bf A}=B\left(
-\frac{q(rx+qy)}{r^2+q^2},\frac{r(rx+qy)}{r^2+q^2}\right) \;,
\end{equation}
which reflects the geometry induced by the electric field. Within this gauge,
the Hamiltonian becomes
\begin{eqnarray}
\label{b7}
(\widehat{H}\psi)_{l,m}=-\frac{J_x}{2}\left( \exp\left[-i2\pi\alpha \frac{q}{N}(rl+qm)\right] \psi_{l+1,m}  +  h.c. \right) + \nonumber \\
-\frac{J_y}{2}\left( \exp\left[i2\pi\alpha \frac{r}{N}(rl+qm)\right]
\psi_{l,m+1} +h.c. \right) + eaF\frac{(rl+qm)}{\sqrt{N}} \psi_{l,m} \;,
\end{eqnarray}
where $h.c.$ denotes the terms required to render the Hamiltonian Hermitian.

The calculation of the energy spectrum can be equally achieved by either of
two different, yet equivalent methods, which we shall refer to as the method
of rotated coordinate frame, introduced in \cite{Naka95}, and the method of
rotated basis.

\subsection{Rotated coordinate frame}

\begin{figure}
\center
\includegraphics[width=8cm,clip]{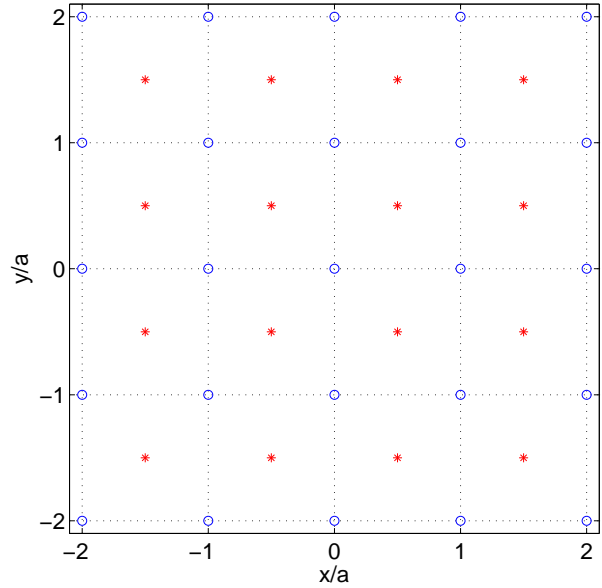}
\caption{Extended lattice for the field direction $(r,q)=(1,1)$, where it
consist of two sublattices.} 
\label{fig3}
\end{figure}

The method of rotated coordinate frame \cite{Naka95} consists of two steps.
The first is to choose the previous gauge, Eq.~(\ref{b2}), for the magnetic
field. The second step is to simplify the Hamiltonian (\ref{b7}) by rotating
coordinates to align the electric field with the vertical axis $\xi$ of a new
coordinate frame $(\eta,\xi)$:
\begin{equation}
\label{b1} \eta=\frac{qx-ry}{\sqrt{r^2+q^2}}  \;,\quad  \xi
=\frac{rx+qy}{\sqrt{r^2+q^2}}  \;.
\end{equation}
In the rotated coordinates, the original lattice sites $(al,am)$, with integer
$l$ and $m$, appear to lie on a sublattice immersed into a new square lattice
of spacing $d$,
\begin{displaymath}
 d=a/\sqrt{N}  \;,
\end{displaymath}
whose sites $(sd,pd)$ can be labeled by pair of integer indexes $(s,p)$. Note
that this new lattice actually consists of $N$ independent sublattices, only
one of which coincides with the original lattice, see Fig.~\ref{fig3}.
Explicitly, the mapping of the original lattice points into the new, extended
lattice is
\begin{displaymath}
(l,m) \rightarrow (ql-rm,rl+qm):=(s,p),
\end{displaymath}
while the inverse mapping reads
\begin{displaymath}
(s,p) \rightarrow \frac{1}{N}(qp-rs,qs-rp):=(l,m).
\end{displaymath}
%

Letting now $\phi_{s,p}$ denote the wave-function amplitude at site $(sd,pd)$
in the rotated frame of reference $(\eta,\xi)$, and using the fact that
$(rl+qm) = p$, we can write the Hamiltonian action as
\begin{equation}
\label{b4b} (\widehat{H}\phi)_{s,p} = -\frac{J_x}{2}\left(e^{-i2\pi\alpha q  p
/ N} \phi_{s+q,p+r}  +  h.c. \right) -\frac{J_y}{2}\left(e^{ i2\pi\alpha r   p
/ N} \phi_{s-r,p+q} + h.c. \right) +edFp \phi_{s,p}  \;.
\end{equation}
Observe that, coherently with Eq.~(\ref{b7}), the $N$ sub-lattices described
above are uncoupled, and that the shift $s \rightarrow s+N$ is an invariant
transformation. Nevertheless, it is convenient to solve the eigenvalue problem
for all sublattices simultaneously. We therefore consider the stationary
Schr\"odinger equation for the complex amplitudes $\phi$,
$(\widehat{H}\phi)_{s,p} = E \phi_{s,p}$. Following \cite{Naka95} we use the
plane wave basis
\begin{displaymath}
\phi_{s,p}=\frac{e^{id\kappa s}}{\sqrt{L}} b_p(\kappa) \;,
\end{displaymath}
where $L$ eventually tends to infinity. In so doing, the quasi-momentum
$\kappa$ belongs to the interval $[0,\frac{2\pi}{a} \sqrt{N})$. We finally arrive
at the following equation, where we have put $\theta = 2\pi\alpha  N^{-1}$:
\begin{equation}
\label{b6} -\frac{J_x}{2}\left(e^{-i \theta q p+iq d\kappa} b_{p+r}  +
e^{i \theta q (p-r)-iq d\kappa} b_{p-r}  \right) -\frac{J_y}{2}\left(e^{i
\theta  r p -i r d\kappa} b_{p+q} + e^{-i \theta  r (p-q) +i r d\kappa} b_{p-q} \right)
+edFp b_{p}=E b_{p} \;.
\end{equation}
Equation (\ref{3}) in Sec.~\ref{sec2}(a) is the particular case $(r,q)=(0,1)$ of this
more general equation.  It is also possible to show (see Appendix) that in the
semiclassical approach this equation yields the classical Hamiltonian
(\ref{a1}), after applying a canonical transformation.

\subsection{Rotated basis}

In this method we do not introduce sub-lattices  but diagonalize the Hamiltonian
by changing the basis. One begins with the tight-binding Hamiltonian in the gauge
(\ref{b2}) as before, that is, Eq.~(\ref{b7}). One then imposes periodicity on
the lattice space in the following way: let $K$ be an even integer, and let
us identify lattice points $(l,m)$ and $(l',m')$ if and only if $l-l'$ is an
integer multiple of $Kq$ and $m-m'$ is an integer multiple of $Kr$. The
Hamiltonian (\ref{b7}) typically does {\em not} satisfy this periodicity.
Nonetheless observe that by letting $K$ tend to infinity the period can be
made arbitrarily large, following the standard approach. At finite $K$ the
lattice is then composed of $K^2 q r$ distinct points.

Next, we define new lattice period
\begin{displaymath}
\tilde{d}=a\sqrt{N}  \;,
\end{displaymath}
and an adimensional quasi-momentum $k=\tilde{d}\tilde{\kappa}$, that can take the
discrete values
\begin{eqnarray}
\label{kappa1}
 k = 2 \pi \frac{j}{K}, \;\; j = 0,\ldots,K-1,
\end{eqnarray}
and a new set of basis functions associated with $k$:
\begin{equation}
\label{b8rev} |\phi_{p,k}^{(\mu)}\rangle=\sum_{n=-K/2+1}^{K/2}
\frac{e^{ik n}}{\sqrt{K}} |p+qn,\mu-rn\rangle  \;.
\end{equation}
This definition is consistent with the imposed periodicity on the lattice, as
can be seen by letting $n \rightarrow n+K$ and observing that the r.h.s. does
not change. Also observe that if $k \neq k'$
\begin{displaymath}
\langle \phi_{p',k'}^{(\mu')} |\phi_{p,k}^{(\mu)}\rangle= 0,
\end{displaymath}
for all choices of $p,p',\mu,\mu'$. Moreover, keeping $k$ fixed,
different functions $|\phi_{p,k}^{(\mu)}\rangle$  and
$|\phi_{p',k}^{(\mu')}\rangle$ overlap if and only if $(p,\mu)$ and
$(p',\mu')$ belong to the same discrete transverse line of direction
$(q,-r)$: {\em i.e.} there exists and integer $j$ such that $(p',\mu') =
(p,\mu) + j (q,-r)$. In this case,
\begin{displaymath}
\label{bg2}
\langle \phi_{p+jq,k}^{(\mu-jr)} |\phi_{p,k}^{(\mu)}\rangle=  e^{i k j}.
\end{displaymath}

The fact that different functions may overlap is clearly a consequence of the
overdetermination of the set $|\phi_{p,k}^{(\mu)}\rangle$, which is made
of $K^3 r q$ elements: the integer variables $p$ and $\mu$ can take all values
from $0$ to $Kq-1$, and from $0$ to $Kr-1$, respectively, while $k$ can
take the $K$ discrete values in Eq. (\ref{kappa1}). To the contrary, as
observed above, the periodic lattice requires a basis of $K^2 r q$ elements
only. Since for each pair $(p,\mu)$, the function
$|\phi_{p,k}^{(\mu)}\rangle$ is a linear combination of $K$ lattice sites
$|l,m\rangle$, it is easy to see that this set can be equally spanned by
keeping $(p,\mu)$ fixed and letting $k$ vary over the $K$ discrete values
in Eq.~(\ref{kappa1}). It is equally easy to find that the full periodic
lattice can then be spanned by choosing $(p,\mu,k)$ in the discrete set
$\cal I$:
\begin{displaymath}
{\cal I} = \{0,\ldots,Kq-1\} \times \{0,\ldots,r-1\} \times \{0,2
\pi \frac{1}{K},\ldots, 2 \pi \frac{K-1}{K} \}.
\end{displaymath}
Remark that in the set $\cal I$ the variable $\mu$ takes on a restricted set
of values, of cardinality $r$. Equivalently, we could have chosen a set
organized in strips parallel to the $m$ axis, but since $r \leq q$, this
choice would have been less convenient.
To sum up, we have introduced the basis for the periodic lattice
%
that, because of the above computations, is composed of orthonormal functions.

Let us now compute the matrix elements of $\widehat{H}$ over this basis. 
The fundamental point is that $\widehat{H}$ does not couple functions with different 
$k$. In fact, when computing the non-diagonal couplings, the product $rl+qm$ appears, that is constant on the transverse lines defined above. The sum over $n$ of the
phases stemming from Eq.~(\ref{b8rev}) is then zero, unless $k =
k'$. The same result also holds for the diagonal term $(F_x l + F_y m)$,
since $(F_x, F_y)$ is proportional to $(r,q)$, see Eq.~(\ref{field}). Observe
that this is true at any finite value of $K$, independently of the fact that
the Hamiltonian $\widehat{H}$ is, or is not, periodic. Therefore, the spectral problem
of $\widehat{H}$ over the periodic lattice decomposes into the $K$ fibers obtained by
letting $k = 2 \pi \frac{j}{K}$, with $j$ fixed. At this point one is
left with the computation of the matrix elements within a fiber.

Suppose now that $1<r<q$, we shall treat the remaining cases $1=r \leq q$.
separately. Then, by explicit calculation we find that the non-zero matrix
elements of $\widehat{H}$ are the following. The ``horizontal'' coupling leads to the
matrix element
\begin{displaymath}
 \langle \phi_{p+1,k}^{(\mu)} | H^{h}
\phi_{p,k}^{(\mu)}\rangle=  -\frac{J_x}{2} \exp\left(i2\pi\alpha
\frac{q}{N}(rp+q\mu)\right) ,
\end{displaymath}
if $0 \leq p \leq Kq-2$, while for $p = Kq-1$
\begin{displaymath}
\langle \phi_{0,k}^{(\mu)} | H^{h}
\phi_{Kq-1,k}^{(\mu)}\rangle=  -\frac{J_x}{2}
\exp\left(i2\pi\alpha \frac{q}{N}(r(Kq-1)+q\mu)\right) .
\end{displaymath}
The ``vertical'' coupling gives
\begin{displaymath}
 \langle \phi_{p,k}^{(\mu+1)} | H^{v}
\phi_{p,k}^{(\mu)}\rangle=  -\frac{J_y}{2} \exp\left(-i2\pi\alpha
\frac{r}{N}(rp+q\mu)\right) ,
\end{displaymath}
if $0 \leq \mu \leq r-2$, while for $\mu = r -1$
\begin{displaymath}
\langle \phi_{p+q,k}^{(0)} | H^{v}
\phi_{p,k}^{(r-1)}\rangle=  -\frac{J_y}{2} \exp\left(-i2\pi\alpha
\frac{r}{N}(rp+q(r-1)) \right) e^{ik},
\end{displaymath}
where $p+q$ at l.h.s. is to be understood modulus $Kq$. The above are half of
the required formulae: we must also add the Hermitian conjugate matrix elements.
Finally, the diagonal coupling is
\begin{displaymath}
\langle \phi_{p,k}^{(\mu)} | H^{d}
\phi_{p,k}^{(\mu)}\rangle = \frac{eaF}{\sqrt{N}} (rp + q \mu).
\end{displaymath}

Observe finally that this lattice problem  can be seen as
a combination of $r$ coupled, one dimensional, periodic lattice problems of
size $Kq$. This takes a particularly simple form in the case $(r,q) = (1,q)$.
In fact, we end up with the single, one-dimensional lattice of size $Kq$, on
which the fiber Hamiltonian has matrix elements
\begin{equation}
\label{bg5} \langle \phi_{p',k}^{(0)} | H \phi_{p,k}^{(0)}\rangle=
-\frac{J_x}{2} \left( e^{i\theta qp} \delta_{p',p+1} + h.c.  \right) 
-\frac{J_y}{2} \left( e^{-i\theta qp} e^{ik} \delta_{p',p+q} + h.c. \right) 
+ \frac{eaFp}{\sqrt{N}} \delta_{p',p}.
\end{equation}
We are therefore equipped with the analytical tools to compute the energy
spectrum of the system.

The right panel in Fig.~\ref{fig1} displays example of the spectrum for $(r,q)=(1,1)$, which was calculated by either of two methods. Note that spectrum is periodic on the quasimomentum with the period $2\pi/\tilde{d}=(2\pi/a)/\sqrt{N}$. Also note that both spectra in Fig.~\ref{fig1} show the characteristic pattern with straight lines: moreover, the slope of lines is the same in both panels. As it will be shown later on, this slope defines the velocity of quantum particle in the transporting regime.

\subsection{Continuity properties of the spectrum}

\label{secabsc} In the preceding paper \cite{85}, both theoretically and in
the numerical examples, we have mainly considered rational and small values of
$\alpha$, a fact that assured both periodicity of the reduced one-dimensional
Hamiltonian in (\ref{3}) and validity of the semiclassical analysis. In this
section we discuss the spectral properties of the Hamiltonian (\ref{basic})
for the case of a general value of $\alpha$. In this section we consider only rational values of the field
direction $\beta$.

As remarked in the introduction, it is instructive to think of the limiting
case of null magnetic and electric field: the spectrum is absolutely
continuous, composed of a single band. Keeping the electric field null, while
turning on the magnetic field, the one-dimensional Hamiltonian (\ref{3}), [or
equally well the Hamiltonian (\ref{b6}) --being the electric field null, the
two are equivalent], becomes the celebrated Harper Hamiltonian, whose spectrum
has bands for $\alpha$ rational: therefore, the spectrum of the original
two-dimensional Hamiltonians  (\ref{1}) and (\ref{b7}), is also absolutely continuous.
For irrational values of $\alpha$, the spectrum of (\ref{3}),(\ref{b6}) is
absolutely continuous (for $J_y>J_x$), pure point (for $J_y<J_x$), or singular
continuous, in the critical case $J_x=J_y$. In the first case, obviously, the
spectrum of (\ref{1}),(\ref{b7}) is absolutely continuous. Interestingly, the
same is true also in the second case, $J_y<J_x$: by varying the quasi-momentum
associated with the invariant direction $x$ the eigenvalues of (\ref{3}) and
(\ref{b6}), that exist since the spectrum is point, {\em move and draw} the
energy bands of (\ref{1}) and (\ref{b7}). Finally, for $J_x=J_y$, since in
this case the spectrum of (\ref{3}),(\ref{b6}) is the same for almost all
values of $\kappa$, due to ergodicity of the Hamiltonian, the spectrum of
(\ref{1}),(\ref{b7}) is also singular continuous. This is the complete
description in the case of null electric field.

The previous results  are classical in historical sense. They are useful to
understand the non-zero electric field case. Since $\beta$ is rational, we can
still use the Hamiltonians (\ref{b7}),(\ref{b6}). In this
case the spectrum of the one-dimensional Hamiltonians (\ref{b6})  
is pure-point, for any value of $\alpha$ -- whether rational or
irrational, small or large. Therefore, we typically expect the spectrum of
(\ref{b7}) to be absolutely continuous, drawn by the eigenvalues
$E_\nu(\kappa)$. Of course, this is heuristic discussion is not a substitute
for a rigorous proof, but it can give us a hint into the physical phenomenon.
We therefore conjecture that for any value of $\alpha$, and non zero electric
field with a rational direction ratio, the spectrum is absolutely continuous.
Quantum dynamics under these circumstances should reflect the spectral
properties and be ballistic. We shall prove numerically that this is indeed
the case.

\subsection{Strong field limit and perturbation theory}
\label{sec3c}

In the strong field limit we can find analytic approximations of the spectrum
by perturbation theory based on Eq.~(\ref{b6}). These results are independent
of the rationality of $\alpha$ and of its amplitude. The case $\beta=0$ has
been described in Sec.~\ref{sec2a}. The next simplest case is $\beta=1$,
{\em i.e.} $(r,q)=(1,1)$. In this case, it is convenient to rewrite Eq.~(\ref{b6}) 
in the form
\begin {displaymath}
 -[V(p;\kappa) b_{p+1} + V^*(p-1;\kappa) b_{p-1}] + edFp b_p=E(\kappa) b_p \;,
\end{displaymath}
where $V(p;\kappa)=\left( J_x e^{-i\pi\alpha p} e^{id\kappa}+ J_y
e^{i\pi\alpha p} e^{-id\kappa}\right)/2$. Similar to the case $(r,q)=(0,1)$
analyzed in Sec.~\ref{sec2a}, the unperturbed spectrum of the system consists
of flat bands separated by the Stark energy,  i.e., $E^0_\nu(\kappa)=edF\nu$.
However, unlike the case $(r,q)=(0,1)$ the first order correction to this
spectrum vanishes. The second order correction is given by
\begin {displaymath}
\Delta E_\nu=\frac{|V(\nu-1;\kappa)|^2}{edF} -
\frac{|V(\nu;\kappa)|^2}{edF}
=\frac{J_xJ_y}{2edF}[\cos(2\pi\alpha(\nu-1)-2d\kappa) -
\cos(2\pi\alpha\nu-2d\kappa)].
\end{displaymath}
This equation proves that the band widths decrease as $1/F$ when $F$
increases. This is a special case of a general perturbation theory result. To
treat all other cases $(r,q)$, notice that according to Eq.~(\ref{b6}), the
Hamiltonian operator can be written as $H = H_0 + V$, with
\begin{equation}
\label{b6x0}
 H_0 = \sum_p edF p \; { | p\rangle \langle p| },
\end{equation}
\begin{equation}
\label{b6x1} V = \sum_p \Phi(p,q,\kappa) {| p\rangle \langle p+r| }+
\Phi^*(p-r,q,\kappa) {| p\rangle \langle p-r| } + \Theta(p,r,\kappa) {|
p\rangle \langle p+q| }+ \Theta^*(p-q,r,\kappa) {| p\rangle \langle p-q|} ,
\end{equation}
where we have put
\begin{equation}
\label{b6x2} \Phi(p,q,\kappa) = -\frac{J_x}{2} \; e^{-i \theta q p} e^{iq
d\kappa}, \;\;
\Theta(p,r,\kappa) = -\frac{J_y}{2} \; e^{i \theta  r p } e^{-i r d\kappa} .
\end{equation}
Two properties of this Hamiltonian are immediately noticed. Firstly, as noted
above, the unperturbed spectrum is equally spaced. Secondly, the perturbation
$V$ only couples unperturbed states (here, Kronecker deltas at site $p$) with
quantum numbers differing by either $q$ of $r$. For short, denote by
$V_{p,p'}$ the matrix elements of the perturbation $V$ in the unperturbed
basis.

When $q>r \geq 1$ it is then immediate that the diagonal terms of the
perturbation $V$ vanish. Proceed next to consider second order perturbation
theory: it is composed of terms of the kind $|V_{p,p+j}|^2/(E^0_p-E^0_{p+j})$,
where $E^0_p$ is the unperturbed spectrum. In our case, $j$ can only take the
values $\pm q$, $\pm r$. Yet, for any of these $j$, the energy difference
$(E^0_p-E^0_{p+j})= -edFj$ is the opposite of that of the term $-j$ (the
unperturbed levels are equidistant), so that the two contributions cancel
exactly, and second order perturbation theory yields a null result.

The last observation has also a bearing on the general, $n$-th term in the
Rayleigh--Schr\"odinger perturbation series for $n \geq 3$, that contains the
second order term as a factor, among other terms. Typically, the $n-th$ term
is composed of sums of products, call them  $\prod$, of $n$ matrix elements of
the perturbation $V$. The first term in such summation is always of the kind
\begin{equation}
\label{b6x4}
  \prod = \frac{V_{p,j_1} V_{j_1,j_2} \cdots V_{j_{n-1},p}}
    {(E^0_p-E^0_{j_1})  (E^0_p-E^0_{j_2})  \cdots (E^0_p-E^0_{j_{n-1}})} \;.
\end{equation}
Other terms in the summation giving the $n$-th term contain shorter ``chains''
of products of matrix elements, of length at most $n-1$, appropriately
multiplied among themselves to give order $n$, of course divided by the
related denominators. In some of these, the second order term appears as a
factor.

Notwithstanding this complexity, the particular form of the Hamiltonian
problem (\ref{b6x0},\ref{b6x2}) permits us to derive a simple result. In
fact, the numerator in Eq. (\ref{b6x4}) is null unless the ``path'' $p
\rightarrow j_1\rightarrow j_2 \rightarrow j_{n-1} \rightarrow p$ is composed
of ``allowed'' jumps of size $\pm q$, $\pm r$. It is easy to realize that, in
order for this path to comprise the least number of jumps, it must contain a
positive number, $s$, of steps of length $q$ and a negative number, $t$, of
steps of length $r$ (or the same with opposite signs), so that $s q + t r =
0$. Now, since $q$ and $r$ are relatively prime, the minimal solutions are
$s=r, t = -q$ or $s=-r, t=q$. This implies that the minimal number of steps
must be $n= r+q$. Therefore, when $q>r \geq 1$ the first non-zero term in the
Rayleigh--Schr\"odinger perturbation series for the eigenvalues is of order
$n=p+q$ and is composed of $2 \bino{q}{p+q}$ addenda of the form (\ref{b6x4}).
All other terms that appear formally in the analytical expression for the
$n$-th term contain shorter ``chains'' (and/or the second order term) and are
therefore null.

The complete expression of the leading perturbation term can be explicitly
computed, using Eq. (\ref{b6x2}). Yet, notice that all terms of the form
(\ref{b6x4}) contain the common factor
\begin{equation}
 \Lambda_{r,q}(F) = \frac{(-J_x)^q (-J_y)^r}{ 2^{q+r} F^{q+r-1}},
\label{b6x5}
\end{equation}
and therefore we can write
\begin{equation}
 E_\nu(\kappa) \simeq  E^0_\nu + \Lambda_{r,q}(F) P_n(\kappa),
\label{b6x7}
\end{equation}
where $P_n(\kappa)$ is a trigonometric function of the quasi--momentum
$\kappa$. A consequence of this result is that the rate of ballistic spreading
discussed in Sec.~\ref{sec2a} is strongly suppressed for large electric field
as soon as the vector ${\bf F}$ does not point to a strongly rational
direction $\beta$. Later on, in Sec.~\ref{sec4a}, we confirm this expectation,
as well as the quantitative dynamical estimates that follow from
Eqs.~(\ref{b6x5},\ref{b6x7}).

\section{Wave packet dynamics: numerical techniques}
\label{sec4}

To simulate numerically the quantum evolution of the system we have adopted
two different approaches. In the first, we have used the time-dependent gauge
${\bf A(t)}={\bf A}_0+c(F_xt, F_yt)$, for which the electric  field appears as
a periodic driving of the system. For the static vector potential ${\bf A}_0$,
which is responsible for the magnetic field,  we used the Landau gauge ${\bf
A}_0=B(0,x)$. This leads to the following Schr\"odinger equation with explicit
time-dependence
\begin{displaymath}
i\hbar\dot{\psi}_{l,m}= -\frac{J_x}{2}\left(e^{-i\omega_x t}
\psi_{l+1,m}  +  h.c. \right) -\frac{J_y}{2}\left(e^{-i(2\pi\alpha l+\omega_y
t)}\psi_{l,m+1} + h.c. \right)  \;,
\end{displaymath}
that we have solved employing the standard Runge-Kutta techniques implemented
in Matlab.

In the second approach, implemented in Fortran, we have chosen the time
independent gauge ${\bf A}=B(-y,0)$, so that the Schr\"odinger equation reads
\begin{displaymath}
i\hbar\dot{\psi}_{l,m}= -\frac{J_x}{2}\left(e^{-i2\pi\alpha m}
\psi_{l+1,m}  +  e^{i2\pi\alpha m} \psi_{l-1,m}\right)
-\frac{J_y}{2}\left(\psi_{l,m+1} + \psi_{l,m-1}\right) +ea(F_x l + F_y m)
\psi_{l,m}.
\end{displaymath}
The r.h.s. of equation is in the form of the action of a time-independent
operator on the two-dimensional lattice vector $\psi$. We have computed the
exponential of this operator using the repeated Chebyshev expansion
\cite{ober}, combined with a numerical truncation of the infinite lattice to a
strip along the line of equation $F_x l + F_y m = 0$, that corresponds to the
direction of spreading of the wave--packet. This is effected numerically by
introducing {\em slanted} integer coordinates $l'=l$ and $m'=m+
\mbox{int}(\frac{F_y}{F_x} l)$. Observe that the directions of increase of
$l'$ and $m'$ are no more orthogonal. Yet, this provides a convenient
re-labelling of lattice sites to enforce the strip truncation mentioned above.
We always checked that the wave--packet projection at the boundaries of large
$|l'|$ and $|m'|$ is smaller than a very low threshold, at any time during the
evolution.

The initial conditions, obviously common to both approaches, consisted of
two--dimensional Gaussian wave--packets, centered at the origin of the
lattice, of adjustable widths: $\psi_{l,m} \sim \exp(-C_x l^2 -C_y m^2)$.
To simulate an incoherent wave packet, all
components $\psi_{l,m}$ have been multiplied by statistically independent
random phases $e^{i \vartheta_{l,m}}$, and an average of quantum amplitudes
over different realizations of the initial packet has been performed.

Finally, to simulate the dynamics of transporting states we have first
constructed them by using the appropriate gauge for the magnetic field and
then we have applied a unitary transformation to translate them into the fixed
Landau gauge. For example, for the transporting states (\ref{6}), which were
constructed using the gauge ${\bf A}_0=B(-y,0)$, the unitary transformation
reads $\psi_{l,m}(t=0) = \exp(-i2\pi\alpha lm) \Psi_{l,m}$.

\section{Wave packet dynamics: Semiclassical region}
\label{secsemevo}

Let us start our analysis of the system dynamics from the semiclassical region
$|\alpha| \ll 1$. It is instructive to have a pictorial look at four significant
cases: we select two values of the field intensity, the first weak, $F=0.2$ and
the second strong, $F=0.5$. We combine these with two values of the orientation
of the electric field: a rational value, $\beta = 2/3$ and a strongly
irrational one, $\beta = (\sqrt{5}-1)/2$.
\begin{figure}
\center
\includegraphics[width=10cm,angle=270,clip]{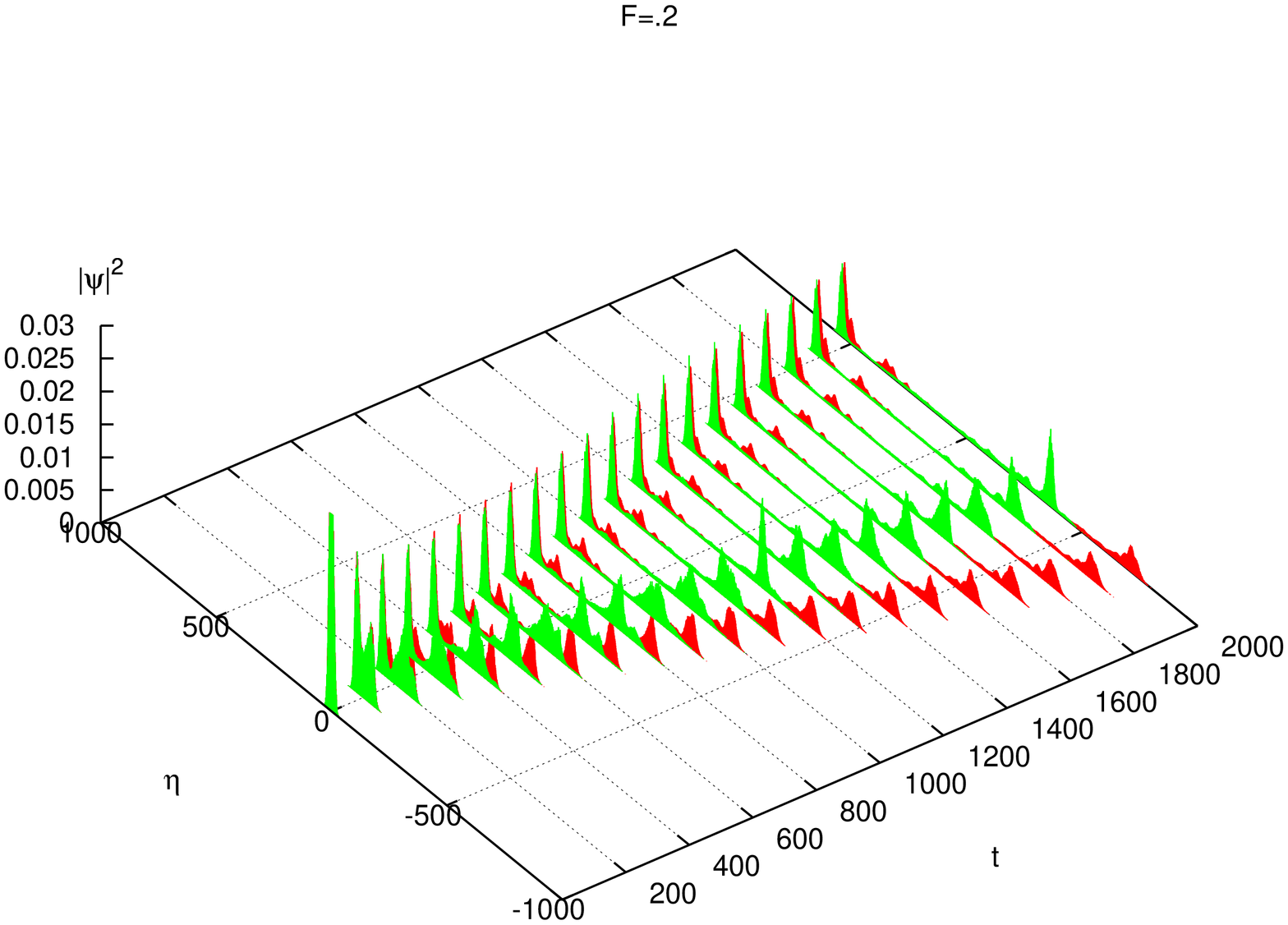}
\includegraphics[width=10cm,angle=270,clip]{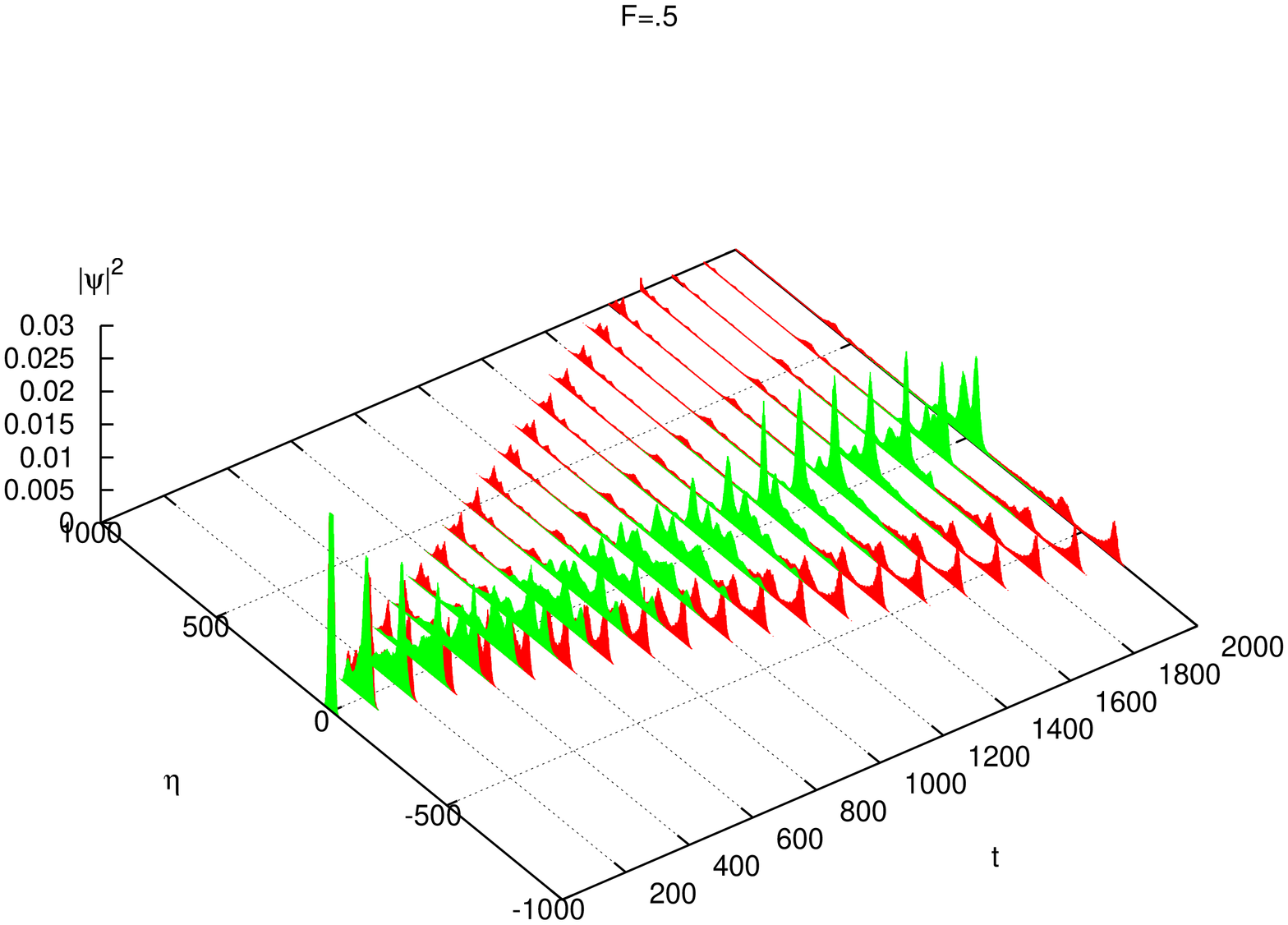}
\caption{(Color online) Wave-packet amplitudes $|\psi(\eta,t)|^2$ versus
$\eta$ and time, averaged over 12 realizations of the initial incoherent
wave-function, for $J_x=J_y=1$, $\alpha = 1/10$ and two electric field
orientations: $\beta = 2/3$ (red curve) and  $\beta = (\sqrt{5}-1)/2$ (green
curve). Electric field amplitude is $F=0.2$ (top) and $F=0.5$ (bottom).}
\label{fig-packp2}
\end{figure}

Since the wave-packet spreads in the direction of the $\eta$ coordinate of the
rotated frame of reference in Eq.~(\ref{b1}), we find convenient to adopt this
coordinate system. Moreover, since we observe that the wave--packet is
localized in the $\xi$ direction, we compute the wave-packet projected
amplitude $|\psi(\eta,t)|^2 = \sum_\xi |\psi(\eta,\xi,t)|^2$.

In Fig.~\ref{fig-packp2} we display $|\psi(\eta,t)|^2$
after averaging this quantity over a number of different realizations of the
random phases in the initial gaussian wave-packet. In the first case, Fig.
\ref{fig-packp2}(a), the electric field amplitude is $F=0.2$, We observe for both
orientation of the field sub-packets moving at constant speed, in the positive
and negative $\eta$ directions. The situation changes radically in Fig.
\ref{fig-packp2}(b), drawn on the same scale, now for $F=0.5$. Here, we find
sub-packets travelling in both directions only in the case of rational
orientation of the field, while in the irrational case we observe localization
of the motion also in the $\eta$ direction.

To the contrary, the projected distributions on the $\xi$ direction (not
reported here), initialized to a Gaussian, settle to a shape that is still
approximately Gaussian, whose width naturally depends on the amplitude of the
electric field: a clear sign of Stark localization in the direction of the
electric field, that does not depend on the orientation of this latter. Let us
now consider separately the weak and strong field cases.

\subsection{Weak field limit and transporting states}
\label{sec3d}

Suppose now that conditions (\ref{basic5}) hold. Then, 
according to Sec.~\ref{sec3a},  
$F<F_{cr}$ and the system has transporting islands.
The quantum-mechanical signature of these islands are straight
lines in the energy spectrum, that are clearly observed in Fig.~\ref{fig1}.
The slopes of these lines coincide with the drift velocity $v^*$ of Eq.~(\ref{basic6}).

Adapting Eq.~(\ref{6})  of Sec.~\ref{sec2c} to the present case of rational
field direction, we can construct a family of localized wave packets, which
move at the drift velocity in the direction orthogonal to the field, without
changing their shape. To do this, we first write these states on the extended
lattice that, we recall, consists of $N$ sublattices:
\begin{equation}
\Phi_{s,p}=\int  g(\kappa) b_p(\kappa) e^{ isd\kappa} {\rm d}\kappa \;.
\end{equation}
In the above, $g(\kappa)$ is an envelope function, that we choose of the form
$g(\kappa) \sim \exp[-C(d\kappa/2\pi)^2].$
%
Then we select from this (generally very large) array only the complex
amplitudes $\Psi_{l,m}$ which sit on the original lattice. We plot these
states in Fig.~\ref{fig5}: the upper row of this figure displays the
transporting states $\Psi_{l,m}$ for $(r,q)=(0,1)$ and $(r,q)=(1,1)$, where we
choose $C=1$.  For this value of $C$ the states are equally localized in both
directions, parallel and orthogonal to the vector ${\bf F}$. If $C$ is
decreased, the states become more extended in the parallel direction and more
localized in the orthogonal direction, see the lower row in the figure. In the
opposite case, {\em i.e.} when $C$ is larger than one, the situation is
obviously reversed. In the limit $C\rightarrow\infty$, when $g(\kappa)$
becomes a Dirac $\delta$-function, the transporting states are extended
Bloch-like waves in the direction orthogonal to the field that carry the
current $v^*$ \cite{85}.

\begin{figure}
\center
\includegraphics[width=10cm,clip]{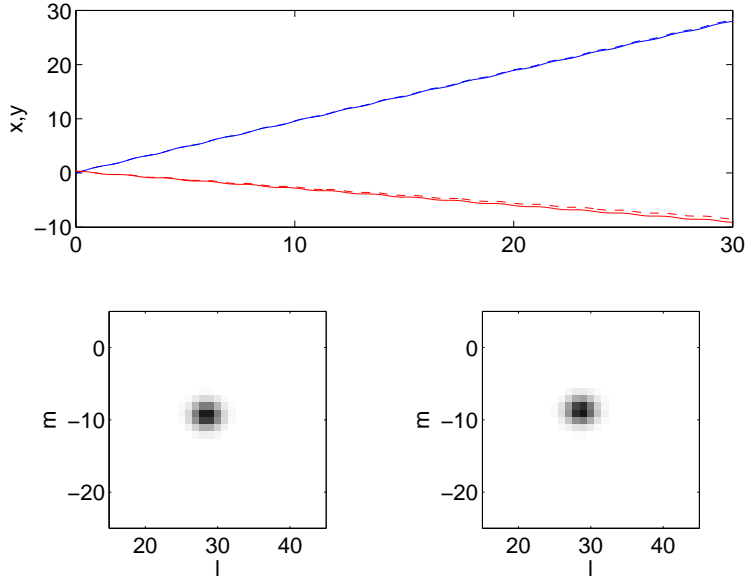}
\caption{(Color online).  The center of gravity of the specially constructed wave packet
$x(t)$ (blue) and $y(t)$ (red) for $eaF/J=0.1$ and $\beta=1/3$ (solid lines)
and $\beta=(\sqrt{5}-1)/4$ (dashed lines) versus time, measured in units of
the tunneling period period $T_J = 2 \pi J/\hbar$ which, in the units used, is
equal to $2 \pi$. The lower panels show the wave packets at the end of
numerical simulations, $t=60 \pi$, depicted as grey tone images (black
maximum, linear intensity scale), for these two cases.} \label{fig9}
\end{figure}

Figure~\ref{fig9} depicts the results of numerical simulations for $F=0.1$ and
two orientations, $\beta=1/3$ and $\beta=(\sqrt{5}-1)/4\approx 0.309$. (Observe that 
these values are half of those considered in the previous subsection.) As initial
condition we choose the wave packet shown in the upper-left panel of
Fig.~\ref{fig5}. This packet is the transporting state for the field direction
$(r,q)=(0,1)$. In his family, it is the easiest to construct numerically.
Moreover, when $F$ is small, it has large overlap with transporting states for
nearby directions as well and it can be used to test the transporting regime
for arbitrary field directions. The solid lines in the panel (a) of
Fig.~\ref{fig9} show the wave-packet center of gravity $(x(t),y(t))$ for
$\beta=1/3$. It is seen that the packet moves in the direction orthogonal to
the electric field: $y(t)=-\beta x(t)$, and that the speeds in the two
directions are those predicted by semi-classical analysis. The dashed lines
show the same quantities, now for $\beta=(\sqrt{5}-1)/4$. The lower panel
shows the wave--packets at the final time of numerical simulations, which
appears to be still well focused. This confirms the fact that in this
short--time, semiclassical regime, dynamics is not affected by
commensurability of Bloch frequencies. We comment later on the long-time
regime.

When the initial wave--function is not in the form of a transporting state, we
encounter different dynamical behaviors. If the initial packet overlaps
significantly a transporting state, we observe a comet-like dynamics with the
comet head moving at the drift velocity and the tail extending in the opposite
direction. Finally, for a generic initial state with small overlap with the
transporting state, the wave-packet dynamics is an asymmetric ballistic
spreading whose dispersion increases in time approximately as
\begin{displaymath}
\label{e4}
\sigma(t)\approx v^*t/\sqrt{2} \;.
\end{displaymath}
However, the most prominent feature of the weak field, semiclassical regime (\ref{basic5}) is that it is insensitive to the rational versus irrational nature of the orientation $\beta$. {\em i.e.} to the
commensurability of the Bloch frequencies. This conclusion is consistent with
the semiclassical analysis.

\subsection{Strong field limit}
\label{sec4a}

In the large field limit, $e a F / 2 \pi J > \alpha$, the scaling law (\ref{b6x5},\ref{b6x7}) implies that
dispersion of a wave--packet is inhibited for irrational field directions.
Therefore, the quantum motion can only oscillate in width and position.  On a
relatively short time scale, this is also the case for ``bad'' rationals $r/q$
with $r,q\gg 1$, while ballistic spreading can be detected only for
$\beta=r/q$ with a small denominator.

\begin{figure}
\center
\includegraphics[width=10cm,clip]{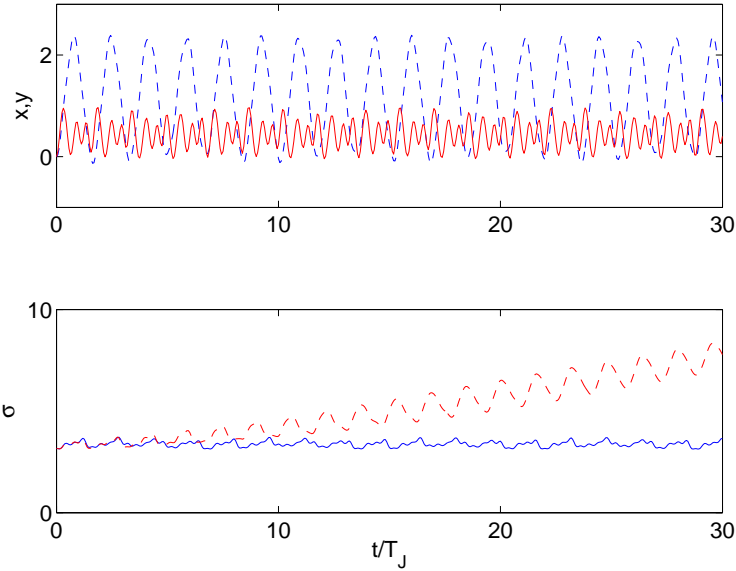}
\caption{ Dynamics of a localized wave packet for $eaF/J=2$, $\alpha=1/10$ and
$\beta=1/3$ (solid line) and $\beta=(\sqrt{5}-1)/4$ (dashed). The lower panel shows the wave-packet
dispersion in these two cases. The upper panel depicts $x(t)$ and $y(t)$ for
$\beta=(\sqrt{5}-1)/4$.}
\label{fig7}
\end{figure}

In the original frame of reference $(x,y)$ the motion can be well described by
the first momenta $x(t)$ and $y(t)$ and by the dispersion
$\sigma^2(t)=a^2 \sum_{l,m}  (l^2+m^2) |\psi_{l,m}(t)|^2 - x^2(t) - y^2(t)\;.$
%
In Fig.~\ref{fig7} we display these data for an initial Gaussian wave packet with 
$F=2$, $\alpha=1/10$ and for rational $\beta=1/3$ and irrational direction
$\beta=(\sqrt{5}-1)/4\approx 0.309$ . The dashed and solid lines in the lower
panel depict the wave-packet dispersion for these two cases, respectively. As expected, a
secular increase of the dispersion is observed only in the rational case
$\beta=1/3$. In the upper panel, dashed and solid lines plot $x(t)$ and $y(t)$ in the case
of the irrational field direction $\beta=(\sqrt{5}-1)/4$. The characteristic
amplitudes and frequencies of oscillations of these quantities are defined by
Bloch oscillations,
\begin{displaymath}
x(t)= \frac{J_x}{2eaF_x} \sin (\omega_x t) \;, \quad y(t)=
\frac{J_y}{2eaF_y} \sin (\omega_y t) \;.
\end{displaymath}
The magnetic field distorts these oscillations, the more the larger the value
of $\alpha$. In particular, for $\alpha$ close to its maximal value 1/2
(without any loss of generality one may consider $|\alpha| \le 1/2$) it
becomes impossible to recognize Bloch oscillations in the time evolution of
the first momenta. Nevertheless, the conclusion that a strong electric field
localizes the quantum particle on a lattice remains valid.

\begin{figure}
\center
\includegraphics[width=9cm,clip,angle=270]{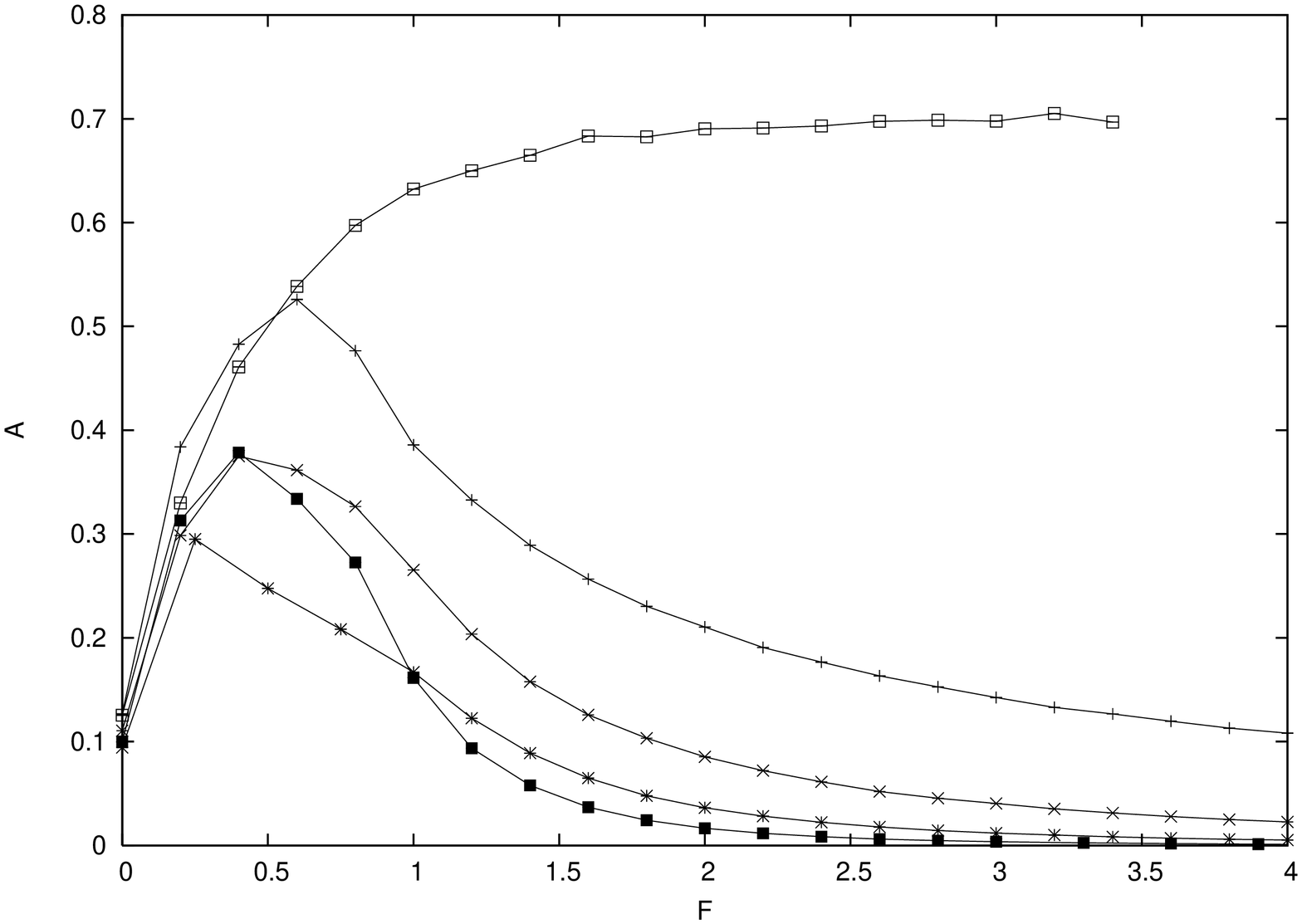}
\includegraphics[width=9cm,clip,angle=270]{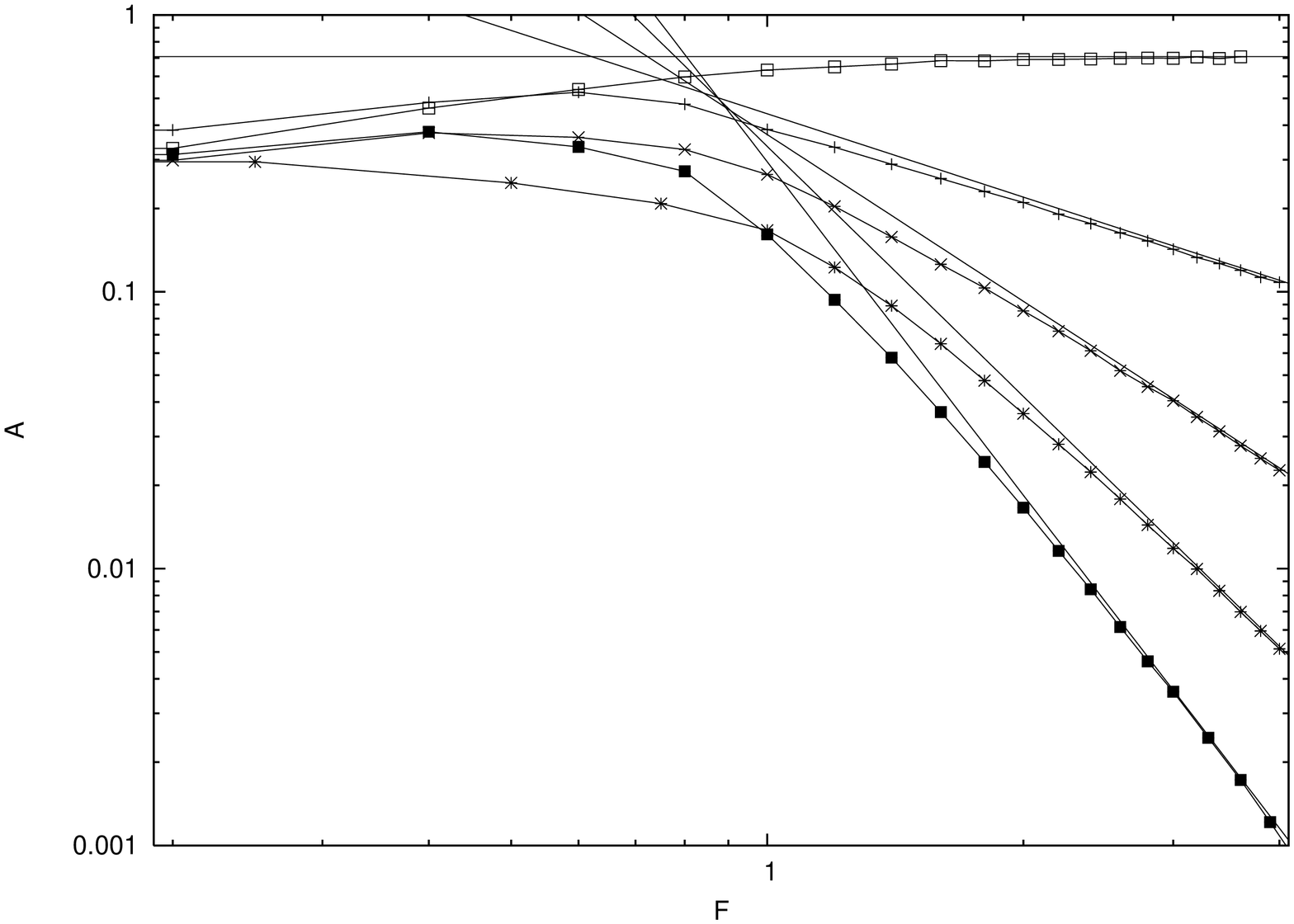}
\caption{Top: Coefficient of the asymptotic growth of the second moment in the
$\eta$ direction versus field amplitude $F$, for $(r,q)=(0,1)$ (open squares);
$(r,q)=(1,1)$ (pluses); $(r,q)=(1,2)$ (crosses); $(r,q)=(1,3)$ (asterisks) and
$(r,q)=(2,3)$ (full squares). Here, $\alpha = 1/10$. Symbols are joined by
lines to guide the eye. Bottom: The same, now displayed in doubly
logarithmic scale. The straight lines are the large field estimates $A \sim F^{1-q-r}$}
\label{fig2lin}
\end{figure}

When delocalization takes place, it is convenient to consider the rotated
coordinate frame $(\eta,\xi)$. This permits to verify numerically the scaling
law (\ref{b6x5},\ref{b6x7}). In fact, we expect that the coefficient $A$ in
the asymptotic growth of the second moment in the $\eta$ direction:
\begin{displaymath}
  M_2(t) :=  a^2 \sum_{\eta}|\psi(\eta,\xi)|^2 \eta^2 \sim A^2 t^2,
\end{displaymath}
be proportional to bandwidth and hence, via Eq.~(\ref{b6x5}), to $F^{1-q-r}$.
In Fig.~\ref{fig2lin} we plot $A$ versus the electric field amplitude $F$,
for five cases of the ratio $\beta$, that takes the rational values zero, one,
one half, one third, and two thirds. Naturally, the second moment in the field
direction $\xi$ is bounded in time. All data sets were computed as averages
over different realizations of incoherent Gaussian wave--packets with the same
initial widths. Values of $A$ were obtained as fits over an asymptotic time
range, extending to a few hundreds (in the units adopted) for $\beta = 0$, and
to tens of thousands, for $\beta = 2/3$.

Bandwidth affects the dynamics in a second way, that might even be more
relevant than the first in laboratory experiments. In fact, the indeterminacy
principle implies that the time required for the dynamics to ``feel'' the
continuous nature of the spectrum is inversely proportional to bandwidth. We
confirmed numerically that the larger the value of $r+q$, the later in time the asymptotic
behavior is achieved: the packet needs more time to unfold and to ``pick up
speed'': at fixed, large $F$, the packet seems to be ``frozen'' for times that
grow exponentially in $q+r-1$, see Eq.~(\ref{basic8}).
\section{Quantum Dynamics in the general case}
\label{secgen}

In the preceding section we have considered small values of $\alpha$, which insures the semiclassical analysis, which obviously describes the quantum motion appropriately only over a finite time
scale. In this section we discuss the extension to the general case.

\subsection{Rational orientation, general Peierls phase}

We have already commented that, for rational orientations $\beta$, the
spectrum should be absolutely continuous for any value of the Peierls phase
$\alpha$. Quantum dynamics under these circumstances should reflect the
spectral properties and be ballistic  even outside the semiclassical
region \cite{remark3}.
\begin{figure}
 \center
\includegraphics[width=10cm,angle=270]{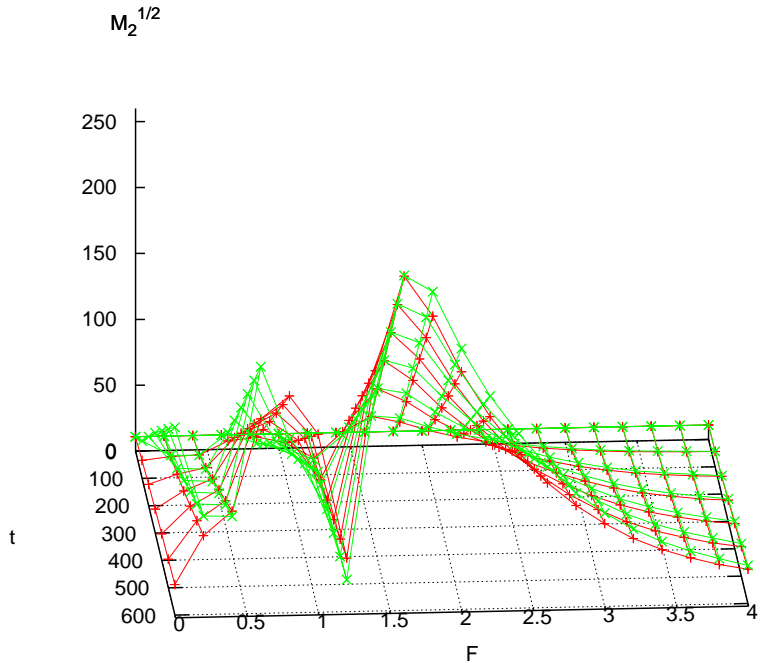}
 \caption{(Color online) Square root of the second moment in
the evolution of an incoherent wave-packet versus time $t$ and field intensity
$F$ for $\alpha=1/3$ (green lines, crosses) and $\alpha=(\sqrt{5}-1)/4$ (red
lines, pluses). The field is directed as $(1,1)$.} \label{figfinx}
\end{figure}

This expectation is confirmed by numerical experiments. In Fig.~\ref{figfinx}
the square root of second moment is shown versus time and field
intensity $F$ for two (large) values of the Peierls phase, $\alpha=1/3$ and
$\alpha=(\sqrt{5}-1)/4 \sim 0.3090$. The orientation is rational: $\beta=1$. As
usual, the incoherent packet is obtained averaging over realization of random
phases. We observe two regions in the plot: as expected, for large $F$ data
are described by perturbation theory, and no difference between the rational
and irrational case is observed. To the contrary, for small $F$ two maxima in
the ballistic speed ({\em i.e.} the slope of the linear growth of $\sqrt{M_2}$)
are observed, with a significant deep between them at around $F=1.25$.
Clearly, this feature is outside the reach of semi-classical analysis.
Moreover, the smaller the value of $F$, the larger become the differences
between the rational and irrational case. At null field, of course, exact
analysis predicts ballistic motion for rational $\alpha$ and anomalous
diffusion, {\em i.e.} quantum intermittency \cite{prl,etna} for irrational
$\alpha$.

To sum up, for rational $\beta$ we find novel behaviors outside the semiclassical region only in the case of small electric fields. Yet, these novel behaviors are at most variations inside a general picture of ballistic
regime.

\subsection{Irrational orientation, general Peierls phase}

Finally, we consider the case of irrational directions of the electric field.
Here, the Hamiltonians (\ref{b4b}) and (\ref{b6}) are not applicable, or
rather they can be used in a sequence of rational
approximations $(r,q)$ to an irrational direction. It seems therefore that in
the perturbative regime the band width (which drives the speed of ballistic
spreading, as well as the time required to start this dynamical regime) is smaller
than any negative power in the field intensity $F$. Yet, it could be a
non-analytic function of this latter. This implies that the wave-packet is
localized for large electric field at any irrational direction $\beta$.
\begin{figure}
 \center
\includegraphics[width=10cm,angle=270]{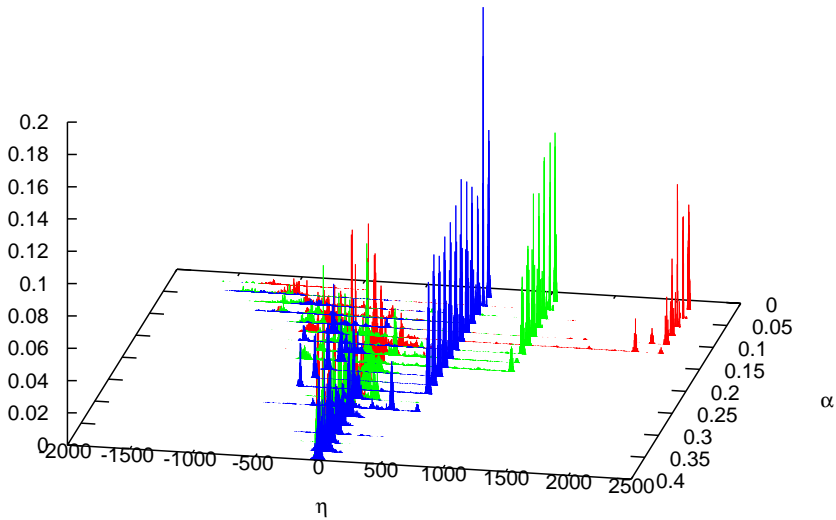}
\caption{(Color online) Final wave-packet at time $t=5000$, versus $\alpha$,
for $v^* = 0.5$ (red), $v^*=0.25$ (green) and $v^*=0.125$ (blue). Here,
$\beta=(\sqrt{5}-1)/2$. Observe the ``classical'' sub-packet traveling in the
positive $\eta$ direction.} \label{figalpha}
\end{figure}

For small $\alpha$ semiclassical analysis (insensitive to the rationality of
the direction) indicates that the motion should be ballistic when $F <
F_{cr}$, {\em i.e.} $F < 2 \pi \alpha J/ea$, a fact confirmed by numerical
experiments, see Fig.~\ref{fig-packp2}. We expect this behavior to hold until
deviations from the semiclassical theory emerge. In fact, the problem of the
spectral type for any electric field intensity should be treated along the
lines of \cite{bellirot}: this analysis should yield the result that the
spectrum is pure point, and quantum motion is localized, albeit the
localization length may be very large, as typically happens in two-dimensional
systems \cite{bellicit}.

Recall that the classical Hamiltonian (\ref{a1}) depends only on the scaled
electric field ${\cal F}$ and its orientation $\beta$. Keeping the direction $\beta$ fixed,
as well as $J=J_x=J_y$, yields a single free classical parameter $\frac{F}{2
\pi \alpha J}$, where $F$ is the amplitude of the electric field. It turns out
that this ratio coincides with the velocity $v^*$ times the dimensional
constant $\frac{\hbar}{Jea^2}$ (that in the units employed in this work takes
the value one): it is therefore interesting to study the quantum dynamics at
fixed $v^*$, while varying the semiclassical parameter $\alpha$. As a
consequence, during this scan, the electric field amplitude scales as $F = 2
\pi \alpha J v^*$. Again, we observe two quite distinct dynamical regions.
Firstly, for values of the classical parameter $v^*$ larger than one, the
motion is always localized about the origin, for any non-zero value of
$\alpha$.
\begin{figure}
 \center
\includegraphics[width=10cm,angle=270]{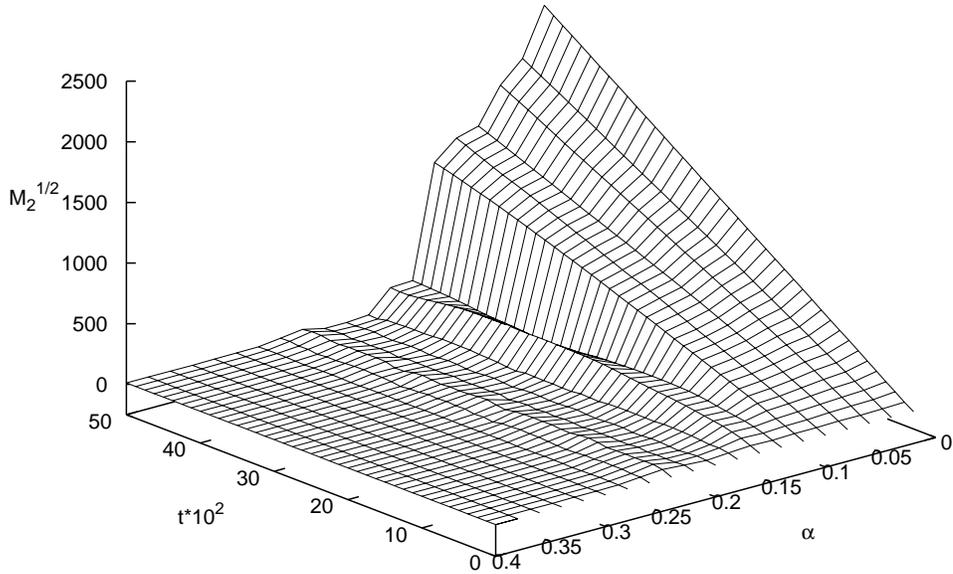}
\caption{Dispersion of the wave--packet versus time and
$\alpha$, for $v^*=0.5$.} \label{figalpha3}
\end{figure}

To the contrary, a rich dynamical behavior is observed for values of $v^*$
smaller than one. In Fig.~\ref{figalpha}  we draw the
$\eta$ projection of the evolved wave-packet amplitude versus $\alpha$, for three
values of the classical parameter $v^*$, and for the irrational value
$\beta=(\sqrt{5}-1)/2$. A coherent sub-packet moving in the positive $\eta$
direction is observed in all cases, roughly independent of the value of
$\alpha$, for small values of this parameter. Clearly, this feature should be
ascribed to the classical dynamics. The region of $\alpha$ in which this
sub-packet is observed diminishes when increasing $v^*$, and disappearance of
the packet is abrupt. For values of $\alpha$ larger than this critical
threshold, the packet remains localized about the origin.

In Fig.~\ref{figalpha3} we plot the dispersion of the evolving wave--packet
as a function of time and $\alpha$, for $v^*=0.5$. The changing dynamical
behavior for increasing $\alpha$ noted in the previous figures is observed
also here: one sees an initial ballistic motion followed by a slower
diffusion and by saturation. These numerical data provide us rough estimates
of the transition time and the localization length. Notice that in this figure
$\alpha$ is proportional to the field intensity $F$: one could therefore
expect a sharp dependence of localization length on the external electric
field, in line with the theoretical remarks presented at the beginning of this
section.

\section{Conclusions}

In our previous work \cite{85} we considered the wave-packet dynamics of a
quantum particle in a square  2D lattice in the presence of a magnetic field
normal to the lattice plane and an electric field, which was aligned along one
of prime axis of the lattice (for definiteness, the $y$ axis). In this work we have
extended these studies to the case of arbitrary direction of the electric
field vector ${\bf F}$. We have confirmed the conjecture of Ref.~\cite{85}
that, depending on the electric field magnitude, the system has two
qualitatively different dynamical regimes, which we refer to as the strong and
weak field regimes, respectively.

The new analysis has extended the validity of the theory developed in
\cite{85} and established new phenomena. For instance, we have used
semiclassical analysis to predict ballistic delocalization for small electric
field intensities, and arbitrary field directions. Under these conditions, we have used spectral
theory to construct non spreading wave--packets traveling at constant speed,
at least for finite times.

In fact, for weak electric field the wave-packet dynamics is governed by the
cyclotron dynamics. This means that the packet moves in the  direction
orthogonal to ${\bf F}$ at the drift velocity $v^*$, which is proportional to
$F$, in close analogy with the problem of a charged particle in a free space
subjected to crossing electric and magnetic fields. However, the presence of
the lattice restricts this behavior to a subspace of initial conditions
discussed in Sec.~\ref{sec3d}. For generic initial conditions we have found
that the packet typically splits into several packets moving in the orthogonal
direction with different velocities, both positive and negative.

In the strong field regime the wave-packet dynamics is governed by the Bloch
dynamics of a quantum particle in a 2D lattice. For null magnetic field, in
these Bloch oscillations the packet oscillates near its initial position.
The obvious exception to this oscillatory
behavior occurs when the vector ${\bf F}$ points to the $y$ direction. Here
the packet spreads ballistically in the  direction orthogonal to the field at
a rate defined by the hopping matrix element. A finite magnetic field
`generalizes' this exception to the cases where the vector ${\bf F}$ points to
a rational direction, i.e., $\beta=r/q$ with $r,q$ being co-prime numbers.
However, now the rate of ballistic spreading in the direction orthogonal to
the field is suppressed by a numerical factor proportional to $(1/F)^{(r+q-1)}$.
This functional dependence implies that in practice the wave packet spreading
can be detected only for simple rational directions with a small denominator
$q$.

We have found a critical field magnitude $F_{cr}$, which separates in the
parameter space the above discussed regimes, by generalizing the semiclassical
approach of Ref.~\cite{85}. This results in a strongly nonlinear 1D classical
system with quasi-periodic driving. Surprisingly, this effective system
appears to be completely integrable in spite of the quasi-periodic character
of driving, which is a rather rare instance from the view point of dynamical
system theory.

Finally we discussed the validity of the semiclassical approach.
For irrational directions $\beta$, at fixed time and classical parameter ${\cal F}$, we observe a sharp suppression of ballistic spreading, occurring when $\alpha$ overcomes a certain threshold.  Combined with results for rational directions $\beta$, this fact leads us to the conclusion that the energy spectrum of the system is continuous for rational $\beta$ and pure point for irrational $\beta$. This result holds for any value of the Peierls phase $\alpha$, both rational and irrational.

\section{Appendix: Semiclassical Hamiltonians}

Let us obtain Eq.~(\ref{a1}) in Sec.~\ref{sec3a} as a semiclassical approximation of 
Eq.~(\ref{b6}).
We follow the same lines as in the derivation of Eq. (\ref{eqoned2}) in Sec.~\ref{sec2c}.
As above, when $|\alpha|$ is much less than one, we replace $p$ by the continuous
variable $\xi$ and we introduce the shift operator $\exp(\partial_\xi)$. At
the left hand side of Eq.~(\ref{b6}) one therefore observes the action of the
operator ${\cal I}$:
\begin{displaymath}
{\cal I}= -\frac{J_x}{2} \left(\exp\left[
  - \frac{iar}{\sqrt{N}} i\partial_\xi -2\pi i \alpha \frac{q}{a \sqrt{N}}\xi \right]
   + h.c. \right)
 -\frac{J_y}{2}\left(\exp\left[
  - \frac{iaq}{\sqrt{N}} i\partial_\xi + 2\pi i \alpha \frac{r}{a \sqrt{N}}\xi \right]
  + h.c. \right)
  +e F \xi \;.
\end{displaymath}
Next, let us introduce the operators $\tilde{X}=2\pi\alpha \xi/a$ and
$\tilde{P}=-i a \partial_\xi$, which obey the commutation relation
$[\tilde{X},\tilde{P}]=2\pi i \alpha$. The previous equation becomes
\begin{displaymath}
  {\cal I}= -J_x \cos\left(
  -\frac{r}{\sqrt{N}} \tilde{P} + \frac{q}{\sqrt{N}} \tilde{X}
  \right)
   -J_y \cos\left(
   \frac{q}{\sqrt{N}} \tilde{P} + \frac{r}{\sqrt{N}} \tilde{X}
   \right)
   +\frac{e a F}{2\pi \alpha} \tilde{X}  \;.
\end{displaymath}
Finally, using the canonical transformation
$Y=(-r/\sqrt{N})\tilde{P}+(q/\sqrt{N})\tilde{X}$,
$P=(q/\sqrt{N})\tilde{P}+(r/\sqrt{N})\tilde{X}$ we obtain the classical
Hamiltonian in Eq.~(\ref{a1}).

\vspace{1.0cm}

{\large \bf Acknowledgements}
We thank Jean Bellissard and Italo Guarneri for illuminating comments on the nature of the energy spectrum for irrational $\beta$. Computations for this work have been performed on the CSN4 cluster of INFN in Pisa and the HPC cluster of Siberian Federal University 
in Krasnoyarsk. 
G.M. acknowledges the support of MIUR-PRIN
project {\em Nonlinearity and disorder in classical and quantum transport processes} and
A.K.  acknowledges the support of SB RAS project {\em Dynamics of atomic Bose-Einstein condensates in optical lattices}.


\begin{thebibliography}{10}
\bibitem{remark1}
By varying this ratio further interesting dynamical effects can be found, see, for example, the work \cite{Ketz}, which discusses the case of null electric field.
\bibitem{remark2}
In Ref.~\cite{85} we restricted ourselves by considering $g(\kappa)=1$ if
$\pi(2n-1)/a\le\kappa<\pi(2n+1)/a$ and zero otherwise.
\end{thebibliography}
\end{document}